**Decoding lithium's subtle phase stability with a machine learning force field**

Yiheng Shen[a] and Wei Xie*[a]

[a] Materials Genome Institute, Shanghai University, Shanghai 200444, China.

*Corresponding Author: xiewei@xielab.org

## ABSTRACT

Understanding the phase stability of elemental lithium (Li) is crucial for optimizing its performance in lithium-metal battery anodes, yet this seemingly simple metal exhibits complex polymorphism that requires proper accounting for quantum and anharmonic effects to capture the subtleties in its flat energy landscape. Here we address this challenge by developing an accurate graph neural network-based machine learning force field and performing efficient self-consistent phonon calculations for *bcc*-, *fcc*-, and 9R-Li under near-ambient conditions, incorporating quantum, phonon renormalization and thermal expansion effects. Our results reveal the important role of anharmonicity in determining Li's thermodynamic properties. The free energy differences between these phases, particularly *fcc*- and 9R-Li are found to be only a few meV/atom, explaining the experimental challenges in obtaining phase-pure samples and suggesting a propensity for stacking faults and related defect formation. *fcc*-Li is confirmed as the ground state at zero temperature and pressure, and the predicted *bcc*-*fcc* phase boundary qualitatively matches experimental phase transition lines, despite overestimation of the transition temperature and pressure slope. These findings provide crucial insights into Li's complex polymorphism and establish an effective computational approach for large-scale atomistic simulations of Li in more realistic settings for practical energy storage applications.

## 1. Introduction

Elemental lithium (Li) metal represents the ultimate anode material for next-generation batteries due to its unparalleled theoretical capacity and the lowest electrochemical potential.[1] However, the practical implementation of Li metal anodes has been hindered by challenges including uncontrolled dendrite growth,[2] making fundamental research on elemental Li crucial for developing viable high-energy-density storage solutions. In addition to technical approaches to suppress these issues,[3, 4] a deeper understanding of Li's intrinsic properties, particularly its phase behavior, is fundamental to addressing these challenges. Despite its apparent simplicity as an alkali metal, Li exhibits a surprisingly complex phase diagram under varying pressure and temperature conditions.[5-8] While *bcc*-Li is favored in the ambient environment, upon cooling it undergoes martensitic transformations to two close-packed structures, namely *fcc*-Li and its hexagonal close-packed stacking variant 9R-Li, which has nine atoms per conventional unit cell.[9-11] The precise nature of their thermodynamic stability and the transitions among them have been subjects of ongoing investigation and debate for decades,[11-17] with a recent combined theoretical and experimental study suggesting that *fcc*-Li could be the ground state[18] despite all the difficulties in obtaining it, calling for more accurate computational techniques built upon precise modeling of quantum and anharmonic effects to account for the subtle energetic differences among the competing phases.[19]

Previous first-principles density functional theory (DFT) simulations have explored the energetics of various polymorphs of Li, predicting near degeneracy between *bcc*- and *fcc*-Li.[20] The inclusion of zero-point energy and finite-temperature anharmonic effects further influences the relative stability, e.g. the vibrational entropy drives the transition from close-packed structures to *bcc*-Li upon heating,[21] and nuclear quantum effects significantly influence the phonon spectra and phase stability.[16] However, these accurate DFT-based calculations are computationally demanding, especially for large-scale simulations required for studying finite-temperature properties. To our knowledge, such limitations have led to the exclusion of the 9R-Li from accurate thermodynamic calculations that properly account for quantum and anharmonic effects, due to its complex crystal structure with significantly lower symmetry than the two competing phases of *bcc*- and *fcc*-Li.

The development of accurate and efficient interatomic potentials has been instrumental in enabling large-scale simulations of Li at length and time scales inaccessible to DFT.[22] Recently, machine learning force fields (MLFF),[23-25] in particular those based on equivariant graph neural networks (GNN),[26-28] have emerged as a promising tool for achieving DFT accuracy in interatomic potentials, while retaining computational efficiency,[29, 30] offering the opportunity to investigate thermodynamics of Li with greater accuracy and computational efficiency. Wang *et al.*[29] developed a deep learning potential (DP) for elemental Li and used it to calculate various bulk, defect and surface properties. Phuthi *et al.*[30] developed a GNN-based MLFF in the NequIP architecture for elemental Li that accurately predicted the small difference (about 2 meV/atom from DFT) in the zero-temperature total energies of *fcc*- and *bcc*-Li, and examined elastic properties, adsorption energies and surface diffusion. However, a systematic thermodynamic analysis of Li leveraging the full power of advanced MLFF remains to be conducted.

In this work, we conduct a comprehensive study of the phase stability of Li in its *bcc*, *fcc*, and 9R phases under near-ambient conditions. We develop a state-of-the-art equivariant GNN-based MLFF in the MACE architecture[27, 28] trained on DFT data, and perform self-consistent phonon (SCP) calculations based on it, accounting for nuclear quantum and the anharmonic vibrational effects of both phonon renormalization and thermal expansion. Temperature and pressure dependent Gibbs free energies of the three phases are calculated afterwards based on the effective force constants obtained from SCP, facilitating the investigation of the stability of the three phases across a range of temperatures and pressures with high fidelity. Our results provide insights into the phase stability of Li and offer practical advances in modeling the thermodynamic properties of this vital element.

## 2. Computational methods

Thermodynamic calculations based on SCP are essentially statistical samplings of the configuration space around the structures of the relevant phases. To facilitate such samplings, our first step was to train a unified MLFF in the MACE architecture[27, 28] for *bcc*-, *fcc*- and 9R-Li based on first principles calculations. To provide training data, a total of 225 supercells with pristine, compressed and stretched lattices of the three phases were generated by the generate_phonon_rattled_structures function implemented in the hiPhive package[31] with the rattling temperature set to up to 900 K. DFT calculations were then performed on these supercells for the energy and interatomic forces using the Vienna *ab initio* simulations package (VASP)[32, 33] with projector augmented wave method[34] for core electrons, the PBE functional for exchange-correlation interactions,[35, 36] and proper plane wave basis energy cutoffs and k-point meshes from careful convergence testing (see ESI). Comparison of PBE, PBEsol[37] and SCAN[38] functionals were performed but no meaningful difference was observed (Figure S1), as expected for this alkali metal. 45 out of the 225 supercells were selected as the test set, which were equally distributed among phases and strains, whereas the rest were used as the training and validation set for the MACE force field. More details are provided in Note S1 in the Electronic supplementary information (ESI).[39, 40]

**3. Results and discussion**

To benchmark the MACE force field developed, the parity plot for the total energy and interatomic forces are shown in Fig. 1a and 1b, respectively. The very small mean average error (MAE) for energy (~1 meV/atom) and forces (~10 meV/Å) along with the near unity of the coefficient of determination $R^2$ in both cases demonstrates reliable modeling of the potential energy surface (PES). In addition to statistics, we also validate the MACE force field by calculating the phonon band structures of *bcc*-, *fcc*- and 9R-Li in the harmonic approximation with the finite-displacement approach as implemented in the phonopy package.[41] The excellent agreement between the phonon band structures calculated by DFT-VASP and MACE force field, as shown in Fig. 1c indicates that the MACE force field has taken in the nuances in the configuration space of elemental Li. Furthermore, as shown in Fig. S2 in the ESI, the nearly indistinguishable equations of states for *bcc-*, *fcc-* and 9R-Li as calculated by DFT-VASP and MACE force field also validate the very high accuracy of the MACE force field over a wide range of pressure. Before proceeding, we wish to note that this particular MACE force field was not trained to describe other complex close-packed stacking[42] structures of Li than *fcc* and 9R, but our approach is readily viable for such endeavors in the future when relevant training data are incorporated.

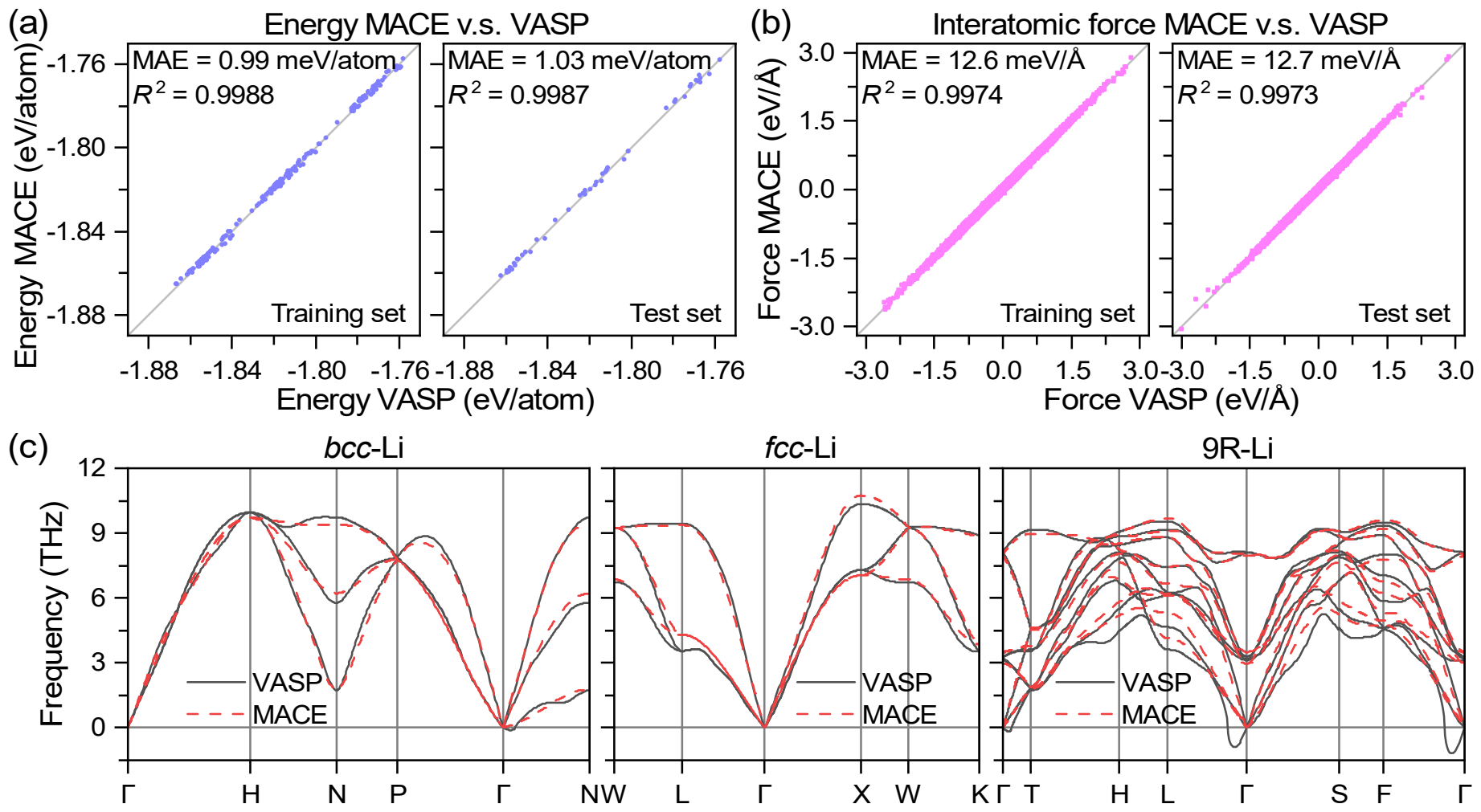

**Fig. 1** Parity plot of (a) total energy and (b) interatomic forces, and (c) harmonic phonon band structures of *bcc*-, *fcc*- and 9R-Li calculated using DFT- VASP and MACE force field. Note that the small imaginary frequency near Γ for the low symmetry 9R-Li phase is a common issue related to the breakdown of acoustic sum rule in DFT functionals.

Using the MACE force field as the energy and force calculator, we then perform SCP calculations for *bcc*-, *fcc*- and 9R-Li at various temperatures and lattice parameters. Lattice parameters are measured in this Communication by strain as calculated with respect to their relaxed values at zero temperature. For 9R-Li which has hexagonal lattice system with two independent lattice parameters $a$ and $c$, a mesh of biaxial strains $\varepsilon_a$ and uniaxial strains $\varepsilon_c$ are considered to account for anisotropic thermal expansion.

Details of the SCP calculations are summarized in Note S2 in the ESI. In essence, the nuclear quantum effects are included by following Bose-Einstein statistics in the sampling of the configuration spaces, phonon renormalization is treated in the SCP calculations each performed at fixed lattice parameters, while thermal expansion is treated by minimizing the Helmholtz free energies from SCP calculations at different lattice parameters following the condition of equilibrium for NPT ensemble, similar to how it is treated in quasi-harmonic approximation. The MACE force field easily accelerated the above calculations by over three orders of magnitude with respect to DFT-VASP calculation, as measured by the wall time of a static calculation of one SCP supercell. See Note S2 of the ESI for more detailed comparison of the computational cost of MACE and DFT-VASP calculations.

Fig. 2 shows the temperature dependent phonon band structures of *bcc*-, *fcc*- and 9R-Li at select lattice parameters, as measured by strain. Fig. S6-S8 provide the results at all calculated lattice parameters. The renormalized phonon modes slightly harden (i.e. the mode frequencies increase) upon heating and profoundly harden upon compression. The dependence of frequency on lattice parameters, which leads to thermal expansion, is found to be more significant than phonon renormalization due to temperature. The two anharmonic effects seem to couple more strongly when Li is subject to compression, and at the strain value of -0.2, soft modes develop for both *bcc*-Li and *fcc*-Li and significant phonon renormalization is found for the two phases. On the other hand, 9R-Li seems to be less sensitive to compression, and remains dynamically stable in the whole range of strains explored in this study. The different reactions to strain are in agreement with the fact that 9R-Li has only been observed under low pressure,[18] as it gains less entropy due to anharmonicity upon increasing pressure than the competing *fcc*-Li phase.

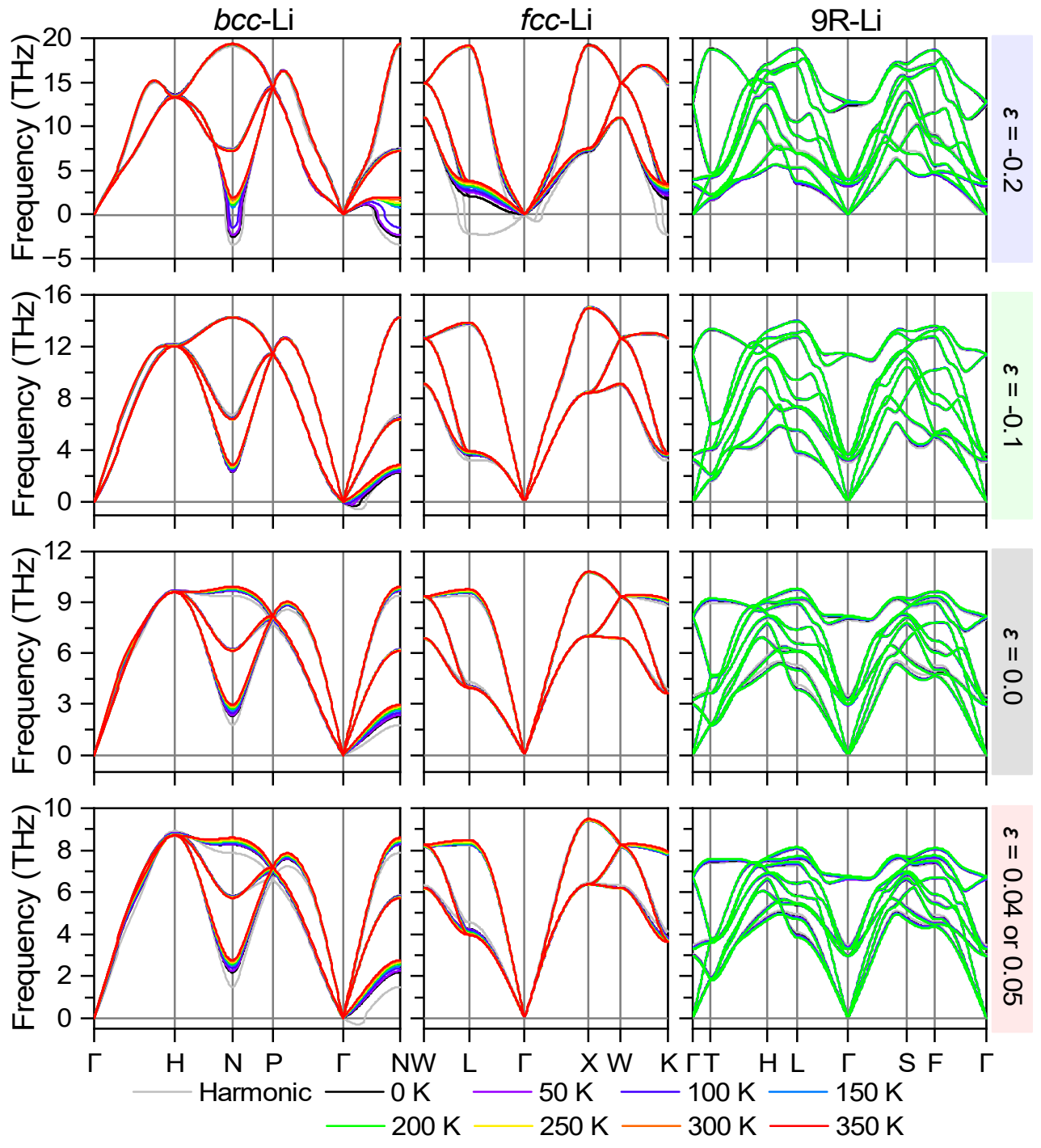

**Fig. 2** Phonon band structures of *bcc*-, *fcc*- and 9R-Li under different isotropic strains. In the last row, the strain value is 0.04 for *bcc*- and *fcc*-Li and 0.05 for 9R-Li. Calculated temperatures range from 0 to 350 K for *bcc*- and *fcc*-Li, and from 0 to 200 K for 9R-Li, with an interval of 50 K.

An important question is then how strong the anharmonicity of Li is. To provide a quantitative estimate, we calculate the anharmonicity measure[43] for the three studied phases. As shown in Fig. 3, the anharmonicity measures of the three phases are comparable to one another. Those of *bcc*- and *fcc*-Li reach 0.3 at 300 K, while that of 9R-Li should still be about the same, as judged from the trend. This result indicates that about 30% of the interatomic forces of elemental Li originate from anharmonic interactions. As a reference, at 300 K the anharmonicity measure of crystalline silicon, a typical material with weak anharmonicity is about 0.2, while that of a strongly anharmonic halide perovskite $KCaF_3$ is about 0.5.[43] This result suggests that anharmonicity in Li is not negligible, which is in accordance with what can be expected from the weak metallic bonding and the light atomic weight of Li, as discussed before,[16, 44] and therefore warrants the careful treatment of both phonon renormalization and thermal expansion in this study.

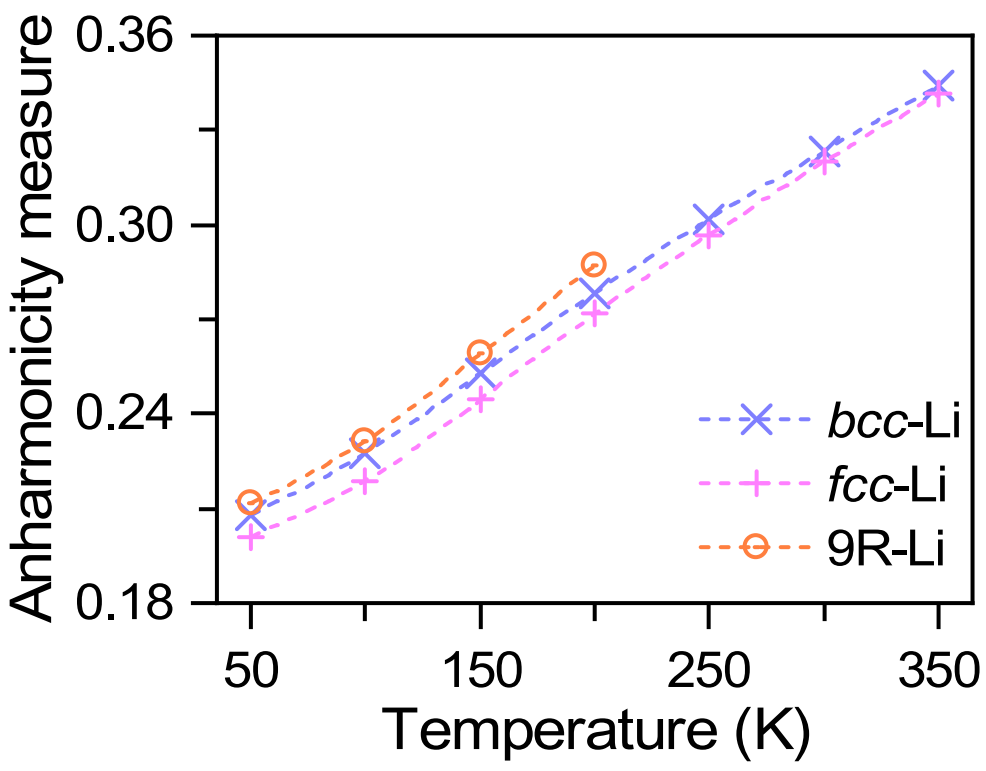

**Fig. 3** Anharmonicity measure of *bcc*-, *fcc*- and 9R-Li as a function of temperature. Analogous results excluding nuclear quantum effects are provided in Fig. S11 in the ESI.

After the lattice parameter and temperature dependent effective force constants are obtained by SCP calculations, we can now calculate the thermodynamic functions of the three phases of Li. The Helmholtz free energy of a phase with lattice **A** at a given temperature $T$, i.e. $F(T, \mathbf{A})$, is obtained as the sum of the vibrational term $F_{\mathrm{vib}}$ calculated based on the effective force constants[45, 46] and the electronic term $F_{\mathrm{elec}}$ calculated based on Fermi−Dirac smearing in DFT by VASP. Details of the $F_{\mathrm{vib}}$ calculations are explained in Note S2 in the ESI. The Gibbs free energy $G$ at given temperature $T$ and pressure $p$, i.e. $G(T, p)$, is then calculated by minimizing the weighted 3rd-order polynomial fit of $F(T, \mathbf{A}) + p{\cdot}\det(\mathbf{A})$ over **A**. We stress that this last step is a non-trivial task, and significant caution needs to be exerted to avoid errors caused by poor fitting, as explained in Note S3 in the ESI.

The predicted temperature- and pressure-dependent Gibbs free energy and equilibrium lattice parameters for *bcc*-, *fcc*- and 9R-Li are summarized in Fig. S9 and S10, respectively. To provide a glimpse into these full results, Fig. 4a shows that the predicted primitive-cell volume of *bcc*-Li coincides with the experimental values at room temperature very well, despite slight underestimation at zero-pressure, which validates the accuracy of our thermodynamic calculations. Fig. 4b shows the pressure-dependent Gibbs free energy at 200 K, in which *fcc*-Li is the ground state throughout the explored range of pressure in agreement with the latest experimental study.[18] It is also worth mentioning that, the Gibbs free energy differences between *bcc*- and *fcc*-Li fall in the few meV/atom range at zero pressure. We recognize that such energy differences are also close to the MAE of the MACE force field, and suspect that the predicted phase diagram has inherited some inaccuracies, as will be discussed below.

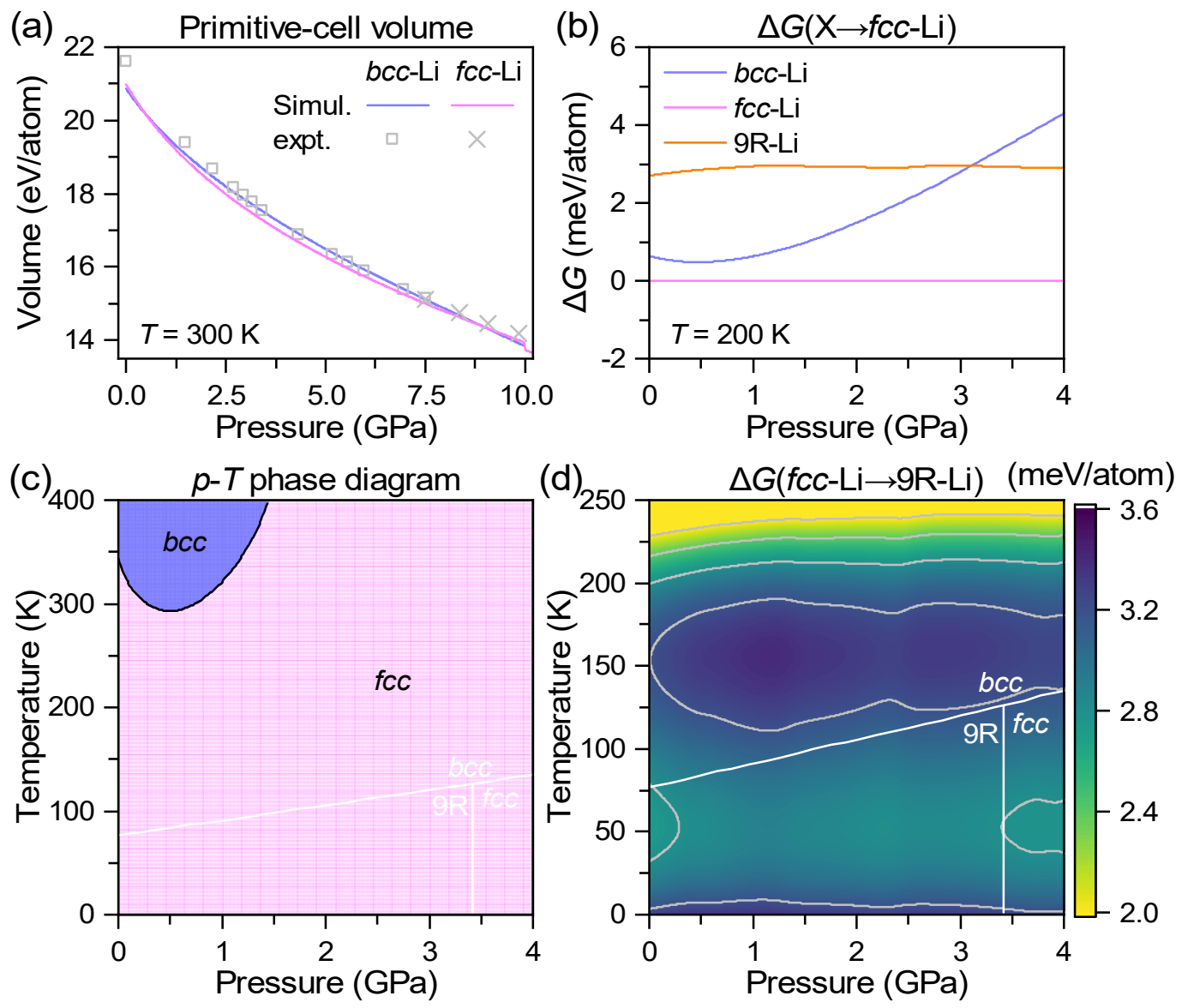

**Fig. 4** Pressure dependent (a) equilibrium primitive-cell volume, (b) Gibbs free energy (reference to *fcc*-Li), (c) phase diagram and (d) Gibbs free energy difference between *fcc*- and 9R-Li. Experimental results in (a) are taken from Ref. 47. White lines in (c) and (d) represent experimental transition lines of $^7$Li, showing that upon cooling at low pressure *bcc*-Li actually transits to metastable 9R-Li instead of *fcc*-Li.[18]

Finally, the continuous *p*-*T* phase diagram of Li is predicted in Fig. 4c. *bcc*-Li is identified as the equilibrium phase at ambient temperature and pressure and its phase boundary with *fcc*-Li shows an upward concave curvature with an extreme point near 0.5 GPa and 300 K. The elevated phase boundary in bcc-Li with respect to pressure above 0.5 GPa qualitatively reproduces the trend of the experimental transition lines between *bcc*- and *fcc*-Li, despite a systematic overestimation of the slope and the critical temperature value (~300 K vs. ~100 K from experiment[18]). The transition lines are only rough estimation of the phase boundary, as they could be significantly affected by various kinetic factors and also depend on the actual experimental paths taken, as discussed extensively in Ref. 18 The seemly large difference between our computationally predicted phase boundaries and the experimental transition lines is also not very surprising given the very flat energy landscape of lithium. In fact, Fig. 4d shows that the Gibbs free energy difference between the two close-packed stacking variants *fcc*- and 9R-Li in the whole 0-4 GPa and 0-250 K range is only 2-3.6 meV/atom. This means that small deviation in Gibbs free energy can lead to the prediction of significantly different phase diagram, as discussed in Note S3 of ESI.

The findings of this study hold significant implications for advancing lithium metal battery technologies, particularly in addressing the persistent challenge of dendrite formation. The near-degeneracy in Gibbs free energy between *bcc*-, *fcc*- and 9R-Li phases in wide ranges of temperature and pressure suggests that even minor perturbations during battery operation—such as localized stress, temperature fluctuations, or interfacial interactions—could trigger phase transitions or coexistence. Such metastability may promote the formation of stacking faults, twin crystal boundaries and other defects in bulk lithium, which act as nucleation sites for heterogeneous lithium deposition and dendrite growth. This underscores the critical need to engineer electrode architectures or electrolyte interfaces that thermodynamically stabilize a single phase under operational conditions. For instance, substrate materials or artificial solid-electrolyte interphases (SEI) designed to epitaxially template *bcc*-Li could suppress defect formation and encourage uniform plating.[48, 49] Furthermore, the pronounced anharmonicity and nuclear quantum effects revealed in this work highlight the necessity of incorporating these factors into computational models to accurately predict lithium's behavior in real-world battery environments. The demonstrated success of the MACE-based approach in capturing subtle phase equilibria at a small computational cost provides a powerful tool for screening materials and operating conditions (e.g., temperature and strain) that favor phase homogeneity through large-scale atomistic simulations of lithium in more realistic settings. By integrating such insights, strategies such as strain-modulated cell designs or thermal management systems could be devised to stabilize desired lithium phases, mitigate stress concentrations, and ultimately enhance the cycling stability and safety of lithium metal batteries. This study thus bridges fundamental understanding of lithium's complex phase behavior to actionable strategies for realizing practical high-energy-density lithium metal anodes.

## 4. Conclusions

In summary, we have performed a detailed computational investigation of the phase stability of *bcc*-, *fcc*-, and 9R-Li under near ambient conditions using a machine learning force field combined with self-consistent phonon theory. The development and validation of an equivariant GNN-based MLFF in the MACE architecture enabled accurate and efficient large-scale thermodynamic calculations for elemental Li. Crucially, our approach rigorously accounts for nuclear quantum effects and the anharmonic effects of phonon renormalization and thermal expansion. Our results reveal the important role of anharmonicity in determining Li's thermodynamic properties, with approximately 30% of interatomic forces arising from anharmonic contributions. The calculated Gibbs free energies reveal exceptionally small energy differences (a few meV/atom) between the competing phases, particularly between *fcc*- and 9R-Li. This near-degeneracy explains the frequent coexistence of phases and the difficulty in preparing phase-pure Li samples, while also suggesting a propensity for stacking faults and related defect formation, which impacts Li metal anode performance. While the predicted *bcc*-*fcc* phase boundary qualitatively agrees with experimental transition lines, the systematic overestimation of the pressure slope and transition temperature underscores the challenges in accurately modeling such a subtle energy landscape and points to areas for future methodological refinement, potentially including more extensive training datasets and/or improved force field architecture. This study demonstrates the power of combining advanced MLFFs with rigorous SCP calculations for probing the intricate phase behavior of complex materials like Li, provides a methodological foundation for future investigations of Li-containing materials and their properties, and suggests directions for the development of better lithium metal battery technologies.

## Acknowledgements

This work was supported by National Science Foundation of China (Grant No. 52003150) and The Program for Young Eastern Scholar at Shanghai Institutions of Higher Learning (Grant No. QD2019006). Y. S. acknowledges the project funded by the Science and Technology Commission of Shanghai Municipality (No. 24CL2901702). This work is supported by Shanghai Technical Service Center of Science and Engineering Computing, Shanghai University.

## References

1. J.-F. Ding, Y.-T. Zhang, R. Xu, R. Zhang, Y. Xiao, S. Zhang, C.-X. Bi, C. Tang, R. Xiang, H. S. Park, Q. Zhang and J.-Q. Huang, *Green Energy Environ.*, 2023, **8**, 1509-1530.
2. M. Li, J. Lu, Z. Chen and K. Amine, *Adv. Mater.*, 2018, **30**, 1800561.
3. A. Chen, X. Zhang and Z. Zhou, *InfoMat*, 2020, **2**, 553-576.
4. Y. Qiu, X. Zhang, Y. Tian and Z. Zhou, *Chinese J. Struc. Chem.*, 2023, **42**, 100118.
5. T. Matsuoka and K. Shimizu, *Nature*, 2009, **458**, 186-189.
6. C. L. Guillaume, E. Gregoryanz, O. Degtyareva, M. I. McMahon, M. Hanfland, S. Evans, M. Guthrie, S. V. Sinogeikin and H. K. Mao, *Nat. Phys.*, 2011, **7**, 211-214.
7. T. Matsuoka, M. Sakata, Y. Nakamoto, K. Takahama, K. Ichimaru, K. Mukai, K. Ohta, N. Hirao, Y. Ohishi and K. Shimizu, *Phys. Rev. B*, 2014, **89**, 144103.
8. M. Frost, J. B. Kim, E. E. McBride, J. R. Peterson, J. S. Smith, P. Sun and S. H. Glenzer, *Phys. Rev. Lett.*, 2019, **123**, 065701.
9. A. D. Zdetsis, *Phys. Rev. B*, 1986, **34**, 7666.
10. H. Smith, *Phys. Rev. Lett.*, 1987, **58**, 1228.
11. V. Vaks, M. Katsnelson, V. Koreshkov, A. Likhtenstein, O. Parfenov, V. Skok, V. Sukhoparov, A. Trefilov and A. Chernyshov, *J. Phys. Condens. Matter*, 1989, **1**, 5319.
12. W. Schwarz and O. Blaschko, *Phys. Rev. Lett.*, 1990, **65**, 3144-3147.
13. P. Staikov, A. Kara and T. Rahman, *J. Phys. Condens. Matter*, 1997, **9**, 2135.
14. O. Blaschko, V. Dmitriev, G. Krexner and P. Tolédano, *Phys. Rev. B*, 1999, **59**, 9095-9112.
15. M. Hanfland, K. Syassen, N. Christensen and D. Novikov, *Nature*, 2000, **408**, 174-178.

16. M. Hutcheon and R. Needs, *Phys. Rev. B*, 2019, **99**, 014111.
17. X. Wang, Z. Wang, P. Gao, C. Zhang, J. Lv, H. Wang, H. Liu, Y. Wang and Y. Ma, *Nat. Commun.*, 2023, **14**, 2924.
18. G. J. Ackland, M. Dunuwille, M. Martinez-Canales, I. Loa, R. Zhang, S. Sinogeikin, W. Cai and S. Deemyad, *Science*, 2017, **356**, 1254-1259.
19. S. H. Taole, H. R. Glyde and R. Taylor, *Phys. Rev. B*, 1978, **18**, 2643-2655.
20. F. Faglioni, B. V. Merinov and W. A. Goddard, III, *J. Phys. Chem. C*, 2016, **120**, 27104-27108.
21. A. Y. Liu, A. A. Quong, J. K. Freericks, E. J. Nicol and E. C. Jones, *Phys. Rev. B*, 1999, **59**, 4028-4035.
22. Z. Qin, R. Wang, S. Li, T. Wen, B. Yin and Z. Wu, *Comput. Mater. Sci.*, 2022, **214**, 111706.
23. A. V. Shapeev, *Multiscale Model. Simul.*, 2016, **14**, 1153-1173.
24. H. Wang, L. F. Zhang, J. Q. Han and W. N. E, *Comput. Phys. Commun.*, 2018, **228**, 178-184.
25. Z. Y. Fan, Y. Z. Wang, P. H. Ying, K. K. Song, J. J. Wang, Y. Wang, Z. Z. Zeng, X. Ke, E. Lindgren, J. M. Rahm, A. J. Gabourie, J. H. Liu, H. K. Dong, J. Y. Wu, C. Yue, Z. Zheng, S. Jian, P. Erhart, Y. J. Su and T. Ala-Nissila, *J. Chem. Phys.*, 2022, **157**, 114801.
26. S. Batzner, A. Musaelian, L. X. Sun, M. Geiger, J. P. Mailoa, M. Kornbluth, N. Molinari, T. E. Smidt and B. Kozinsky, *Nat. Commun.*, 2022, **13**, 2453.
27. I. Batatia, D. P. Kovacs, G. Simm, C. Ortner and G. Csányi, *Adv. Neural Inf. Process. Syst*, 2022, **35**, 11423-11436.
28. I. Batatia, S. Batzner, D. P. Kovács, A. Musaelian, G. N. Simm, R. Drautz, C. Ortner, B. Kozinsky and G. Csányi, *arXiv preprint arXiv:2205.06643*, 2022.
29. H. Wang, T. Li, Y. Yao, X. Liu, W. Zhu, Z. Chen, Z. Li and W. Hu, *Chinese J. Chem. Phys.*, 2023, **36**, 573-581.
30. M. K. Phuthi, A. M. Yao, S. Batzner, A. Musaelian, P. Guan, B. Kozinsky, E. D. Cubuk and V. Viswanathan, *ACS Omega*, 2024, **9**, 10904-10912.
31. F. Eriksson, E. Fransson and P. Erhart, *Adv. Theory Simul.*, 2019, **2**, 1800184.
32. G. Kresse and J. Furthmuller, *Phys. Rev. B*, 1996, **54**, 11169-11186.
33. G. Kresse and J. Furthmuller, *Comput. Mater. Sci.*, 1996, **6**, 15-50.
34. P. E. Blöchl, *Phys. Rev. B*, 1994, **50**, 17953-17979.
35. J. P. Perdew, K. Burke and M. Ernzerhof, *Phys. Rev. Lett.*, 1996, **77**, 3865-3868.
36. J. P. Perdew, M. Ernzerhof and K. Burke, *J. Chem. Phys.*, 1996, **105**, 9982-9985.
37. J. P. Perdew, A. Ruzsinszky, G. I. Csonka, O. A. Vydrov, G. E. Scuseria, L. A. Constantin, X. L. Zhou and K. Burke, *Phys. Rev. Lett.*, 2008, **100**, 136406.
38. J. W. Sun, A. Ruzsinszky and J. P. Perdew, *Phys. Rev. Lett.*, 2015, **115**, 036402.
39. W. S. Morgan, J. J. Jorgensen, B. C. Hess and G. L. W. Hart, *Comput. Mater. Sci.*, 2018, **153**, 424-430.
40. G. L. W. Hart, J. J. Jorgensen, W. S. Morgan and R. W. Forcade, *J. Phys. Commun.*, 2019, **3**, 065009.
41. A. Togo and I. Tanaka, *Scr. Mater.*, 2015, **108**, 1-5.
42. C. H. Loach and G. J. Ackland, *Phys. Rev. Lett.*, 2017, **119**, 205701.
43. F. Knoop, T. A. R. Purcell, M. Scheffler and C. Carbogno, *Phys. Rev. Mater.*, 2020, **4**, 083809.

44. M. Borinaga, U. Aseginolaza, I. Errea, M. Calandra, F. Mauri and A. Bergara, *Phys. Rev. B*, 2017, **96**, 184505.
45. Y. Oba, T. Tadano, R. Akashi and S. Tsuneyuki, *Phys. Rev. Mater.*, 2019, **3**, 033601.
46. T. Tadano and W. A. Saidi, *Phys. Rev. Lett.*, 2022, **129**, 185901.
47. M. Hanfland, I. Loa, K. Syassen, U. Schwarz and K. Takemura, *Solid State Commun.*, 1999, **112**, 123-127.
48. L. Li, S. Li and Y. Lu, *Chem. Commun.*, 2018, **54**, 6648-6661.
49. J. Liu, Z. Bao, Y. Cui, E. J. Dufek, J. B. Goodenough, P. Khalifah, Q. Li, B. Y. Liaw, P. Liu and A. Manthiram, *Nat. Energy*, 2019, **4**, 180-186.

## Note S1. MACE force field training

The training set for the MACE force field was generated as follows. Supercells of *bcc*-, *fcc*- and 9R-Li are generated from their pristine and strained primitive cells. The strain values, supercell matrices and temperature values used in the generation of phonon-rattled structrues are summarized in Table S1. For each strained primitive cell, five phonon-rattled supercells are respectively generated at 300, 600 and 900 K, yielding a total of 15 structures per primitive cell and thus a total of 15 structures per strain value × 5 strain values per system × 3 systems = 225 structures.

**Table S1** Strain values and the supercell matrices for generating rattled structrues of *bcc*-, *fcc*- and 9R-Li, together with the number of atoms in respective primitve cell and supercell ($N_{\mathrm{prim}}$ and $N_{\mathrm{sc}}$).

| System | Strain values | Supercell matrix | $N_{\mathrm{prim}}$ | $N_{\mathrm{sc}}$ |
|---|---|---|---|---|
| *bcc*-Li | −0.03, −0.02, 0.0, 0.02, 0.05 | $\begin{bmatrix} 0 & 6 & 6 \\ 6 & 0 & 6 \\ 6 & 6 & 0 \end{bmatrix}$ | 1 | 432 |
| *fcc*-Li | −0.03, −0.02, 0.0, 0.02, 0.05 | $\begin{bmatrix} -5 & 5 & 5 \\ 5 & -5 & 5 \\ 5 & 5 & -5 \end{bmatrix}$ | 1 | 500 |
| 9R-Li | −0.03, −0.02, 0.0, 0.02, 0.03 | $\begin{bmatrix} -7 & 7 & 0 \\ 4 & 4 & -8 \\ -1 & -1 & -1 \end{bmatrix}$ | 3 | 504 |

First-principles calculations were performed on the rattled structures by using the Vienna *ab initio* simulations package (VASP)[1, 2] based on density functional theory, where the projector augmented wave method[3] was used for describing the interaction between valence electrons and ion cores, and the plane waves with an energy cutoff of 500 eV were used to expand the electron wave functions. The exchange-correlation interaction among the valence electrons is calculated by the PBE functional.[4, 5] The integration over the first Brillouin zone was performed on a 2×2×2 **k**-point grid in the Monkhorst−Pack scheme for calculating energy and interatomic forces in the supercell and a generalized regular **k**-point grid[6, 7] with KSPACING = 0.025 $Å^{-1}$ for calculating electronic entropy and enthalpy in the primitive cell. The convergence thresholds for the total energy during the self-consistent field iteration and the interatomic forces during the geometry optimization were set to $10^{-5}$ eV and $10^{-3}$ eV $Å^{-1}$, respectively.

As shown in Fig. S1a and S1b, KSPACING = 0.025 $Å^{-1}$ and energy cutoff of 500 eV yielded converged energy profile, with denser **k**-point grid (smaller KSPACING value) and higher energy cutoff amending the energy by less than 0.1 meV/atom. The 2×2×2 **k**-point grid for the pristine and strained supercells of *bcc*-, *fcc*- and 9R-Li in our training data aligns with the converged grid density of 0.025. The phase-resolved lattice parameters optimized with different exchange correlation functionals are comparable (Fig. S1c). The converged lattice parameters and energy profiles are summarized in Table S1, which consists with previous studies, e.g., the energy of *fcc*-Li is lower than that of *bcc*-Li by 2 meV/atom.[8]

The MACE force field quantitatively reproduced the energy difference between *bcc*- and *fcc*-Li, as summarized in Table S2, which is not a trivial task since it approaches the RMSE of the force field. In addition, the MACE force field reproduces the energy profile of *bcc*-, *fcc*- and 9R-Li upon compression to atomic volume of about 13 Å$^3$, as shown in Fig. S2, validating its application in high-pressure simulations.

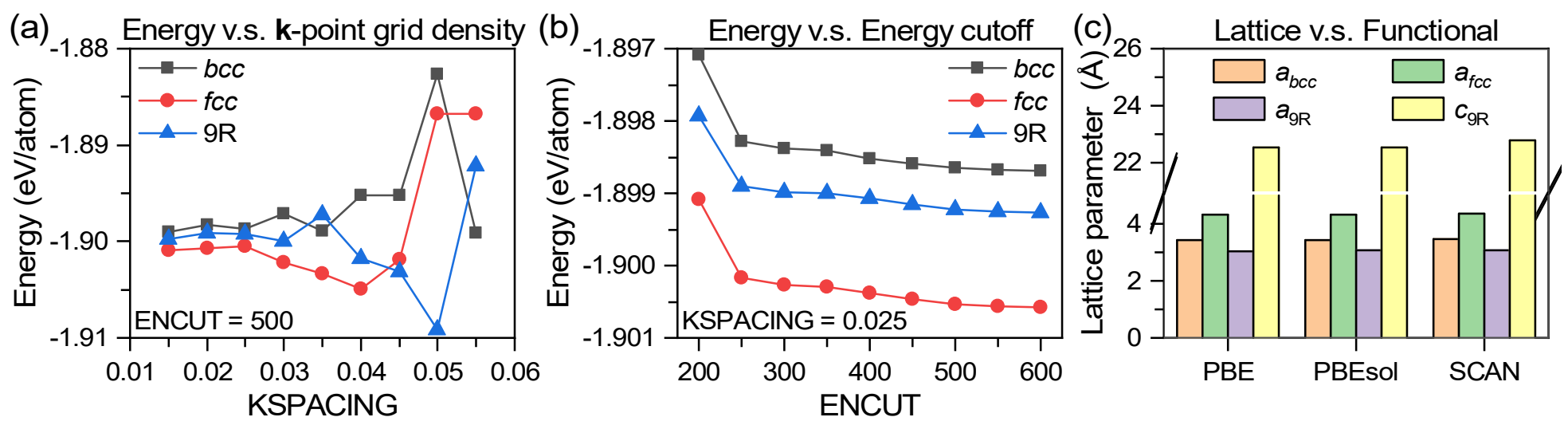

**Fig. S1** Energy over (a) KSPACING parameter for generalized regular **k**-point grid density and (b) energy cutoff. (c) Lattice parameters optimized with different exchange-correlation functionals.

**Table S2** Phase resolved equilibrium lattice parameters $a_{fcc}$, $a_{bcc}$, $a_{9\mathrm{R}}$, $c_{9\mathrm{R}}$ (in Å) and energy difference $\Delta E$ between *bcc*- and *fcc*-Li (in meV/atom) predicted by DFT and MLIPs in this work and previous research. Energy cutoff ($E_{\mathrm{cut}}$, in eV) and k-point grid density ($d$, in Å$^{-1}$) for DFT simulations are provided for reference.

| Model | $a_{fcc}$ | $a_{bcc}$ | $a_{9\mathrm{R}}$ | $c_{9\mathrm{R}}$ | $\Delta E$ | $E_{\mathrm{cut}}$ | $d$ |
|---|---|---|---|---|---|---|---|
| DFT (this work) | 4.330 | 3.440 | 3.061 | 22.52 | 1.9 | 500 | 0.025 |
| MACE (this work) | 4.338 | 3.436 | 3.101 | 22.76 | 2.2 | -- | -- |
| DFT[8] | -- | 3.427 | -- | -- | 2 | 1360 | 0.02 |
| DeepMD[8] | -- | 3.434 | -- | -- | 1 | -- | -- |
| NequIP64[8] | -- | 3.429 | -- | -- | 1 | -- | -- |

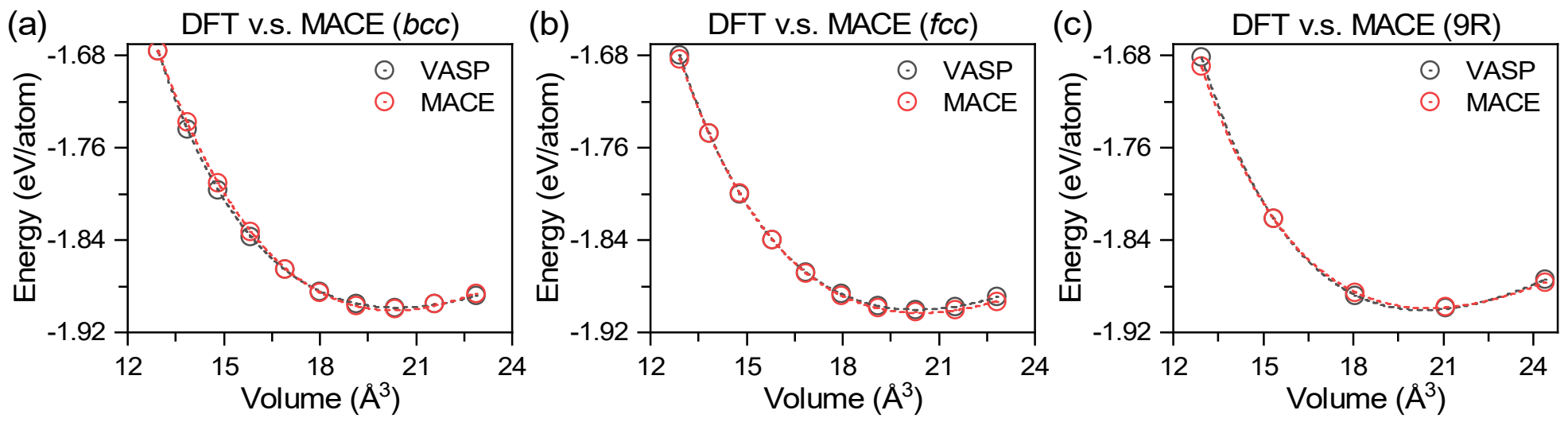

**Fig. S2** Energy of (a) *bcc*-, (b) *fcc*- and (c) 9R-Li upon isotropic change of volume, as calculated by VASP (black) and MACE force field (red).

**Note S2. SCP and thermodynamic calculations**

In each SCP iteration, 100 rattled supercells, whose supercell matrices are identical with those during the generation of training data for the MACE force field, are generated from the latest effective harmonic model $\mathbf{\Phi}_{n-1}$ *via* phonon rattling. The interatomic forces of these rattled supercells are then calculated by the MACE force field. The new effective harmonic model $\mathbf{\Phi}_n$ is obtained from a linear mixing of the least-square fit of the force-displacement relation in the rattled supercells and $\mathbf{\Phi}_{n-1}$ with mixing parameter alpha = 0.1. A total of 30 SCP iterations are performed at each temperature studied. That is, interatomic forces in 30 × 100 = 3000 rattled supercells were calculated to obtain SCP in a given lattice at a single temperature. Evaluating the interatomic forces of a 500-atom supercell of *fcc*-Li required approximately one hour using 64 CPU cores with DFT-VASP, and less than one second using a single NVIDIA V100 GPU with MACE. This represents a three-orders-of-magnitude speedup in wall-clock time, even before considering that buying 64 CPU hours is typically more expensive than one V100 GPU hour from most high-performance computing service providers. A complete self-consistent phonon (SCP) calculation cycle necessitates the evaluation of approximately two million such supercells. This translates to an estimated 500 GPU hours using MACE, compared to approximately 100 million CPU hours using VASP, a cost that is likely prohibitive and perhaps unprecedented.

Vibrational Helmholtz free energy $F_{\text{vib}}$ is calculated from the effective harmonic frequencies $\omega_{\mathbf{q}\nu}$ obtained from SCP by

$$F_{\mathrm{vib}} = U_0 + \frac{1}{2}\sum_{\mathbf{q}\nu}\hbar\omega_{\mathbf{q}\nu} + k_{\mathrm{B}}T\sum_{\mathbf{q}\nu}\ln\left[1-\exp\left(-\frac{\hbar\omega_{\mathbf{q}\nu}}{k_{\mathrm{B}}T}\right)\right],$$

where **q** is the reciprocal coordinate, *v* is the phonon band index, and

$$U_0 = \left\langle V\left(\mathbf{R}_0+\mathbf{x}\right) - \frac{1}{2}\mathbf{x}^{\mathrm{T}}\mathbf{\Phi}_{\mathrm{eff}}^{(2)}\left(T\right)\mathbf{x}\right\rangle$$

is the reference energy, $V(\mathbf{R})$ is the potential energy surface represented by the MACE force field, $\mathbf{\Phi}_{\mathrm{eff}}^{(2)}$ is the effective harmonic model from SCP calculation, **x** is the displacement with respective to the equilibrium configuration $\mathbf{R}_0$, and <·> denotes the ensemble average over the phonon rattling **x** with Bose−Einstein distribution. The summation over **q** is performed on uniformed grids of 26×26×26 for *bcc*-Li, 16×16×16 for *fcc*-Li and 18×18×18 for 9R-Li, so that the their free energy values are respectively converged to < 0.1 meV/atom.

## Note S3. Free energy fitting

The fitting of the Helmholtz free energy $F$ as a function of lattice parameter(s) **A** is not a trivial task. When raw SCP data are calculated at uneven portions of compressive and tensile strained lattice parameters, ordinary least-squares (OLS) fitting provides poor results, as it is sensitive to the number of high-energy raw data points. Take the Helmholtz free energy of *bcc*-Li at $T = 350$ K and 0 GPa as an example, when we successively discard the most compressed raw data point, the fitted curves change dramatically around their minima (Fig. S3b) despite the visual similarity in the entire range (Fig. S3a) and the high $R^2$ values of over 0.999.

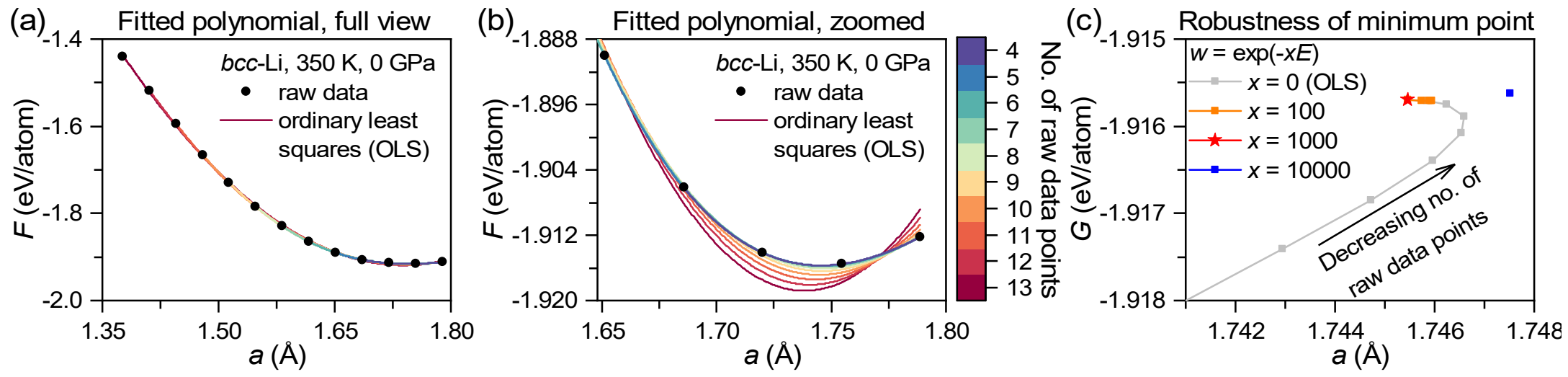

**Fig. S3** Helmholtz free energy $F$ of bcc-Li as a function of lattice parameter a at 350 K and 0 GPa: (a) full view of all 13 raw data points (black dots) from SCP calculations and the fitted curves, and (b) zoomed-in view near the equilibrium lattice and the fitted curves, showing only five visible raw data points. Color of the curves differentiates the number of raw data points fitted to (see the color palate on the right). (c) Evolution of minimum points from 3rd-order polynomials fitted with different weight functions.

To obtain robust fitting results, we performed weight least squares (WLS) fitting instead. That is, the raw data is considered with weight $w(E) = \exp(-xE)$ in the least squares fitting, where $x$ is an empirical parameter and $E$ (in eV/atom) is the sum of Helmholtz free energy $F$ and the volumetric term $pV$, i.e. $E = F + pV$. Proper $x$ value was explored and set respectively to 1000, 1000 and 300 for *bcc*-, *fcc*- and 9R-Li, where smaller values cannot cancel out the data dependency and larger values tend to overly bias the minimum of the fitted curve toward the minimal raw data. For comparison, we compared the results from OLS (left) and WLS (right) fitting in Fig. S4 and S5. One can see that the errors from OLS fitting result in qualitatively wrong prediction of 9R-Li as ground.

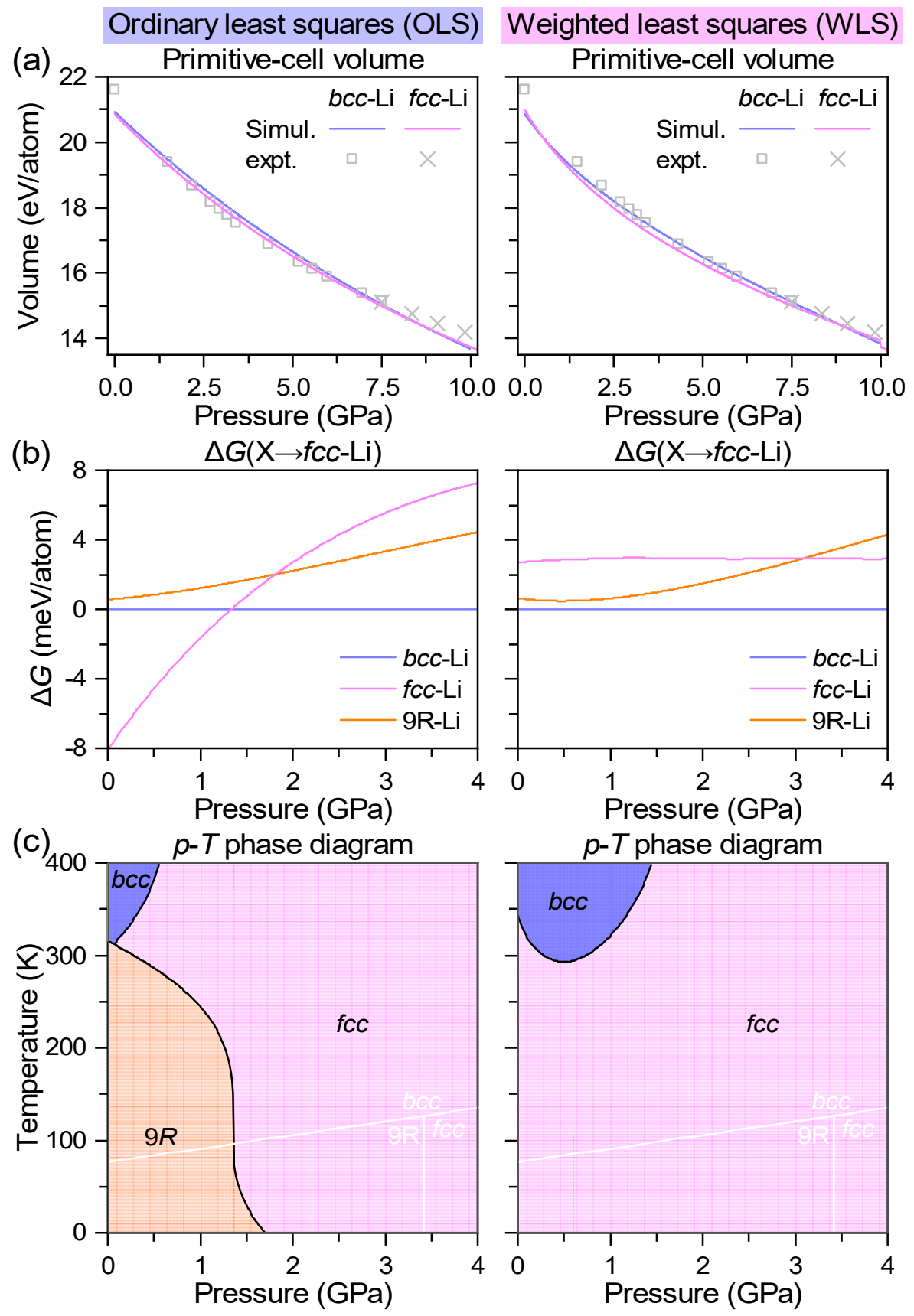

**Fig. S4** Pressure-dependent (a) primitive unit cell volume (at 300 K), (b) Gibbs free energy difference (at 200 K) and (c) phase diagram as predicted from ordinary least squares (OLS) (left) and weighted least squares (WLS) (right) fittings, respectively using raw SCP data.

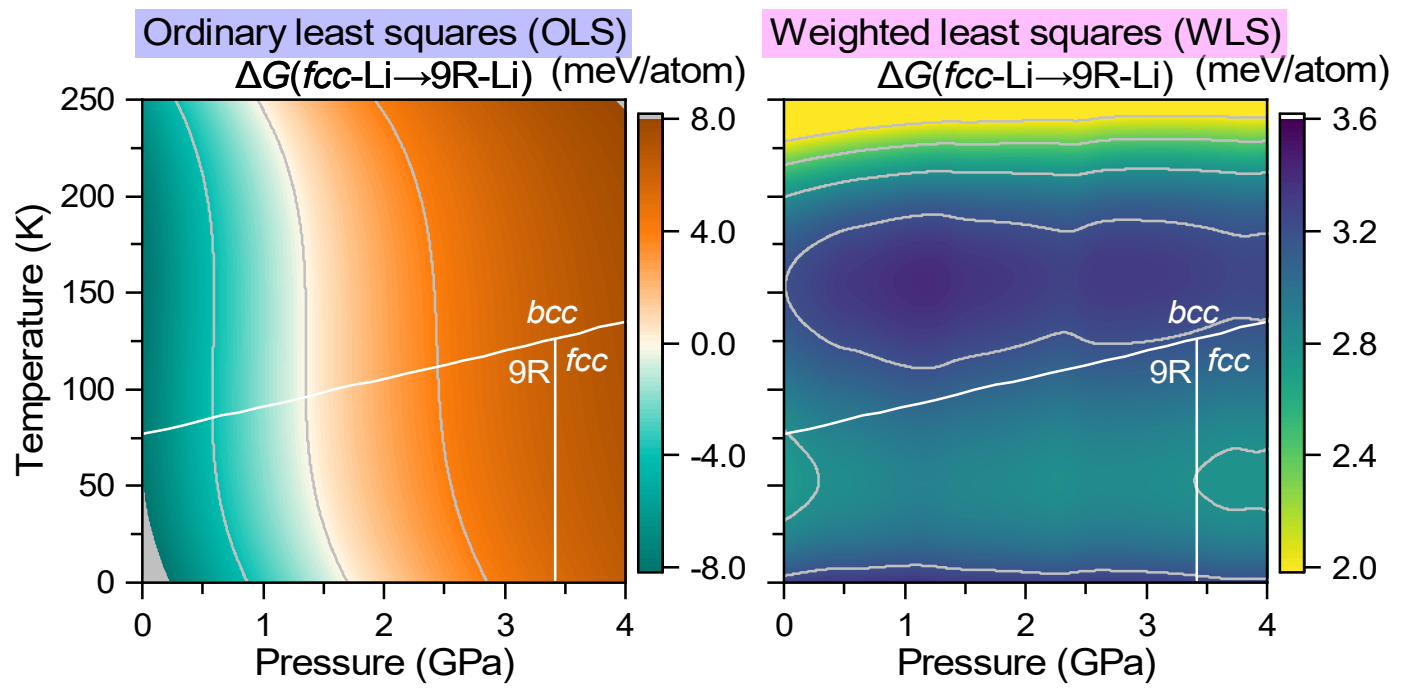

**Fig. S5** Gibbs free energy difference between *fcc*- and 9R-Li calculated from ordinary least squares (OLS) (left) and weighted least squares (WLS) (right) fittings, respectively.

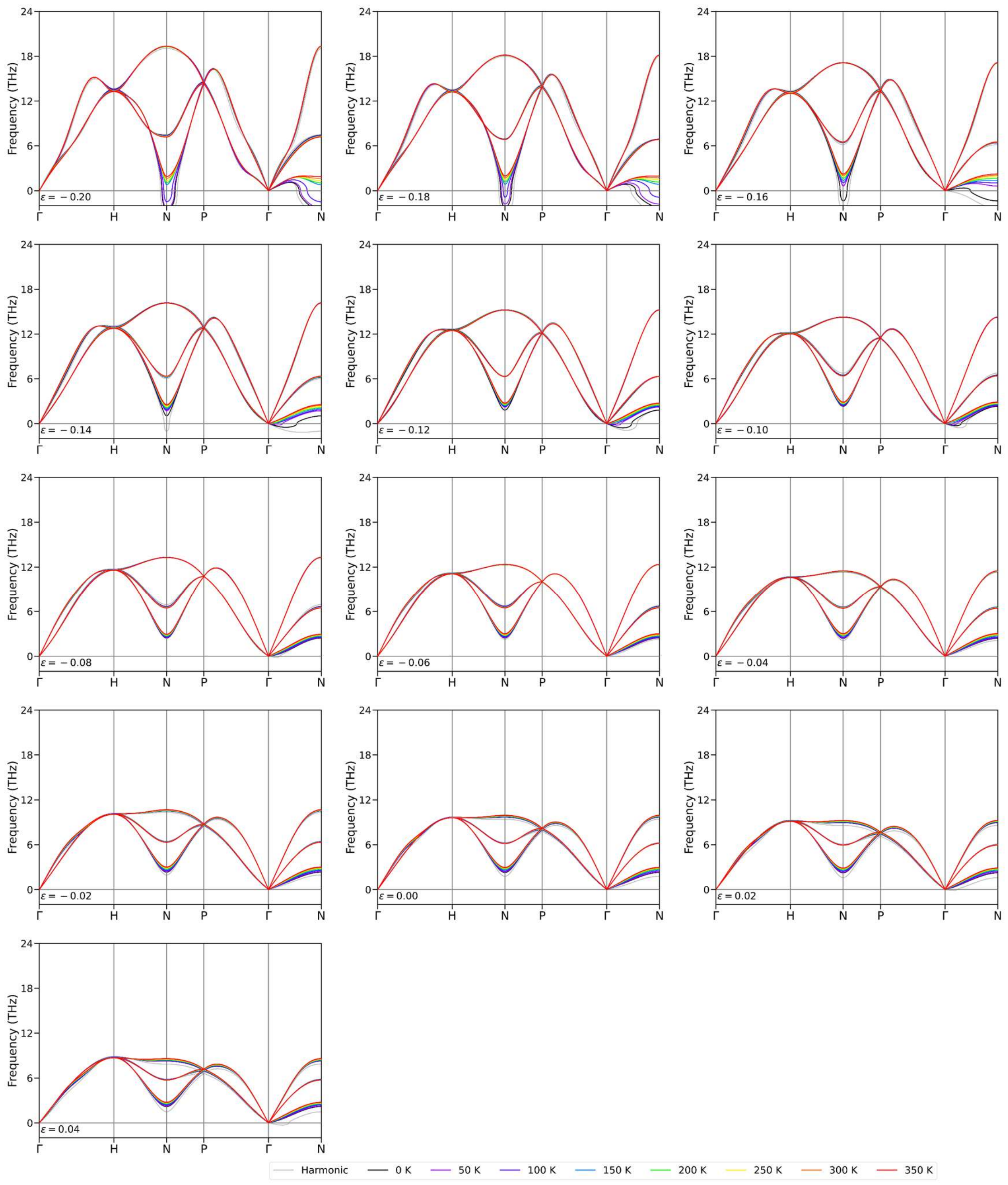

**Fig. S6** Phonon band structures of *bcc*-Li under equiaxial strain $\varepsilon$.

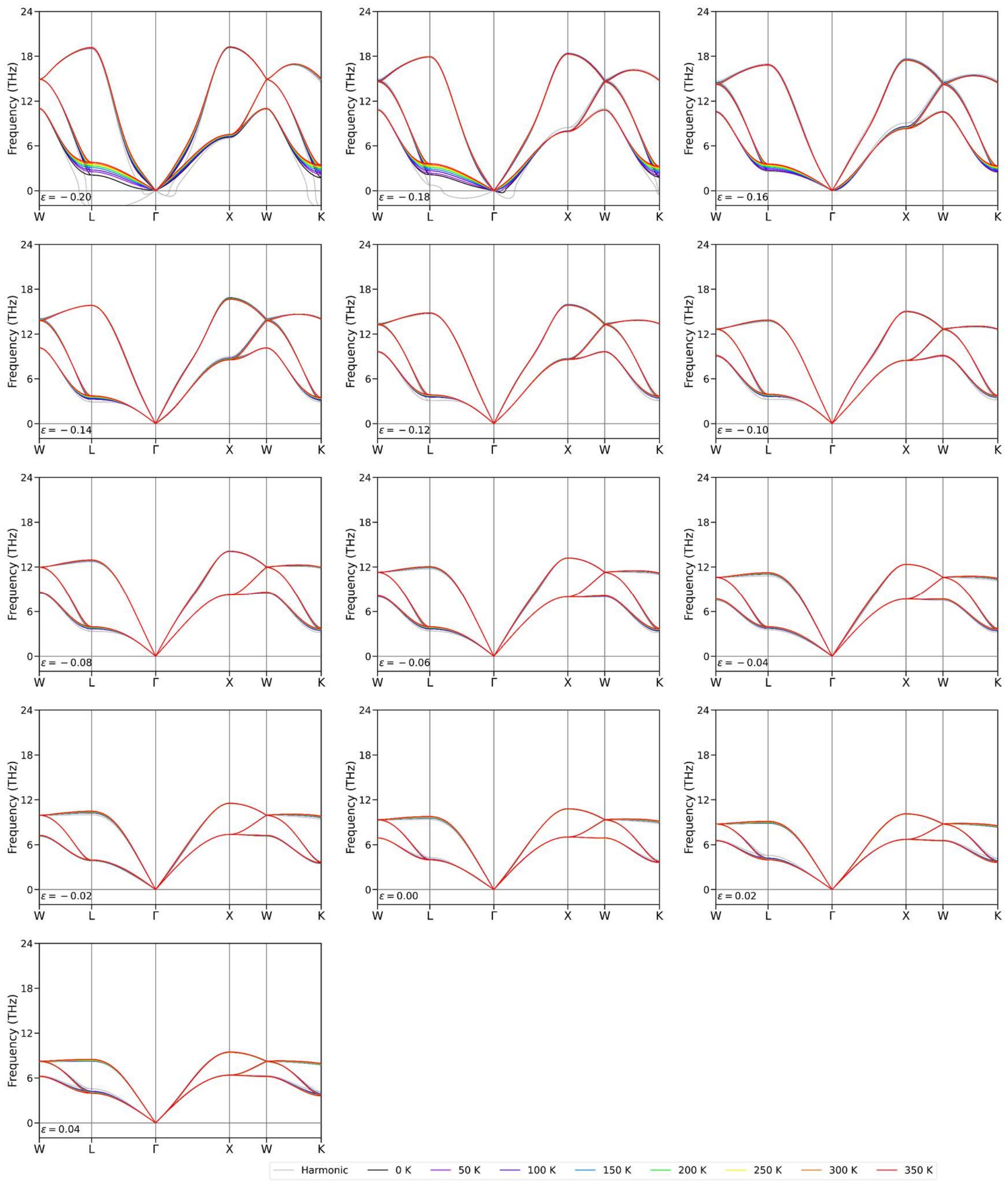

**Fig. S7** Phonon band structures of *fcc*-Li under equiaxial strain $\varepsilon$.

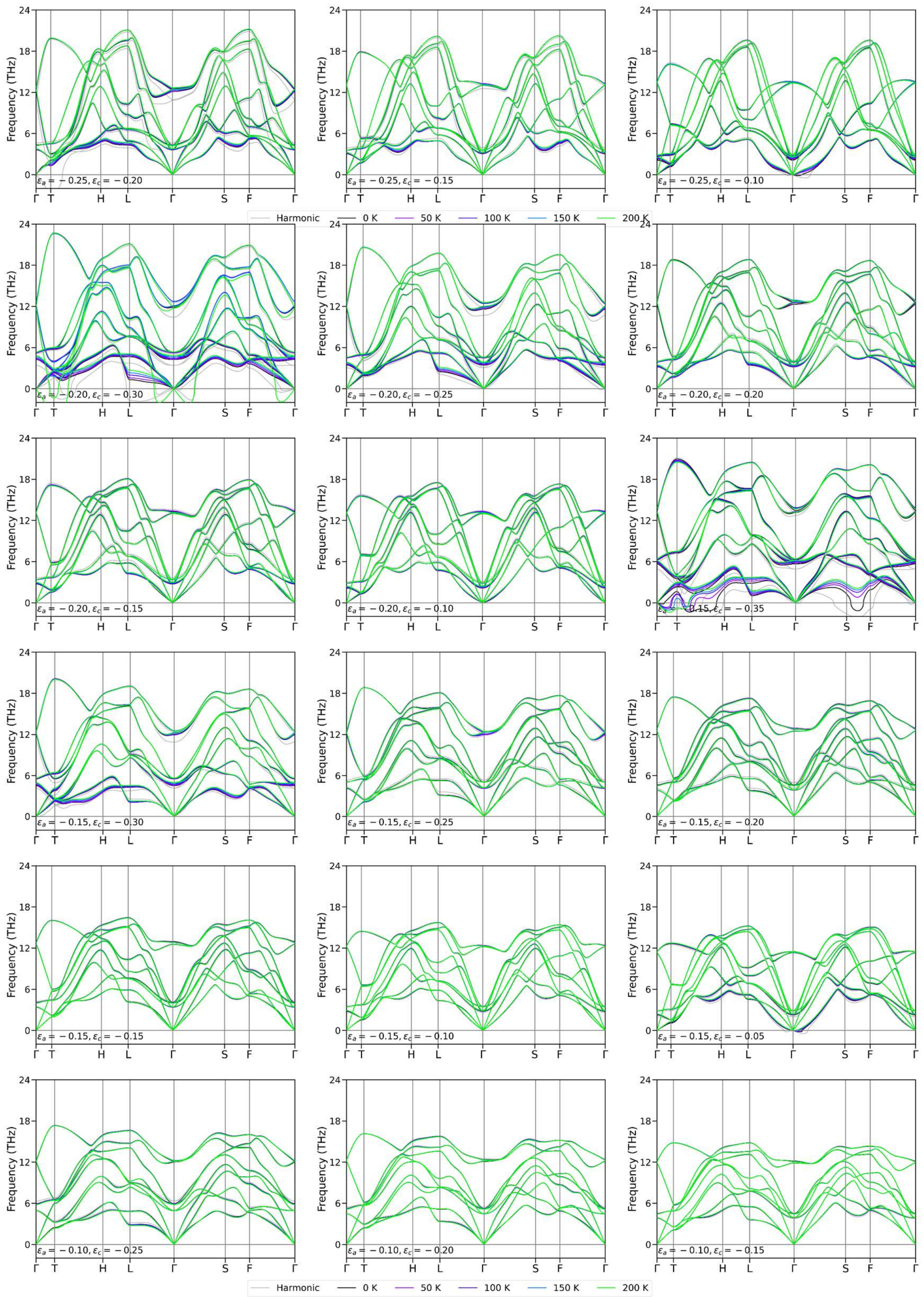

**Fig. S8** Phonon band structures of 9R-Li under biaxial strain $\varepsilon_a$ and uniaxial strain $\varepsilon_c$.

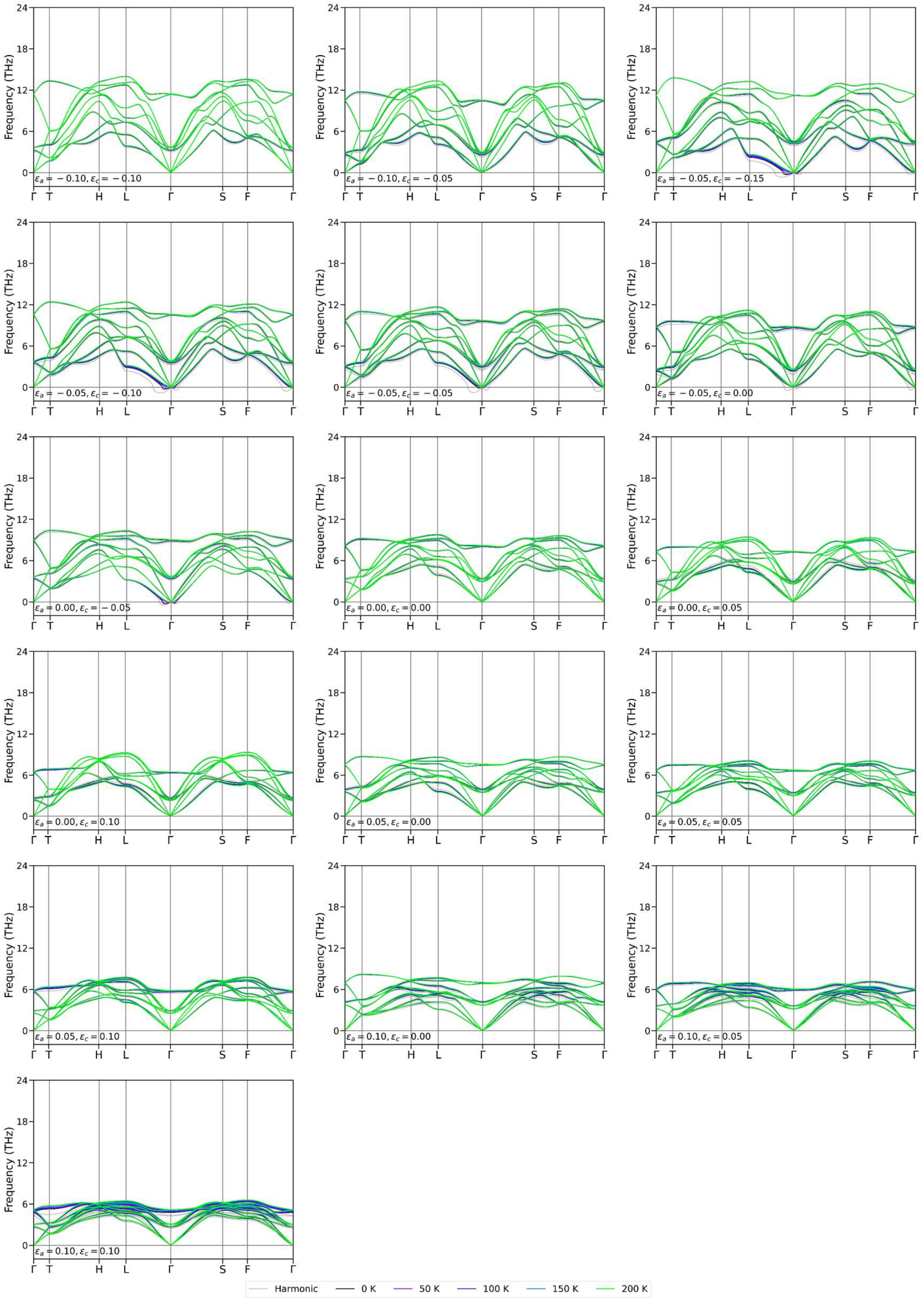

**Fig. S8** (continued) Phonon band structures of 9R-Li under biaxial strain $\varepsilon_a$ and uniaxial strain $\varepsilon_c$.

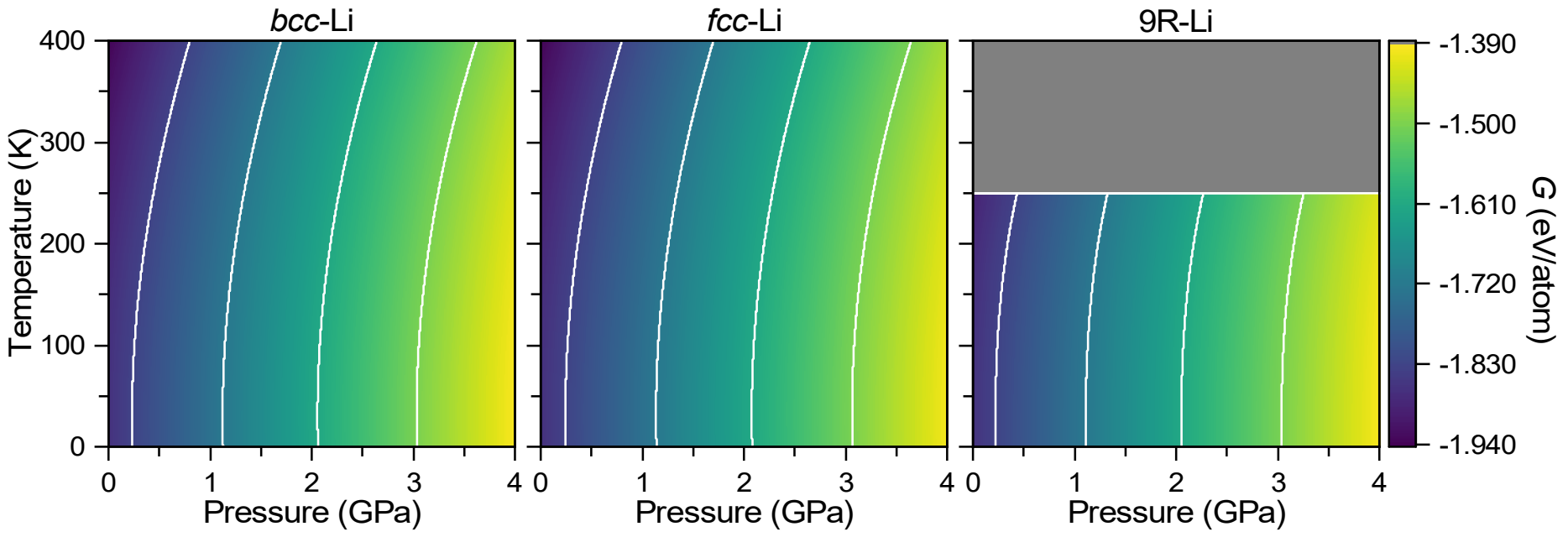

**Fig. S9** Temperature- and pressure-dependent equilibrium Gibbs free energy of *bcc*-, *fcc*- and 9R-Li. Grey region in 9R-Li panel represents missing data to avoid over-extrapolation from calculated results.

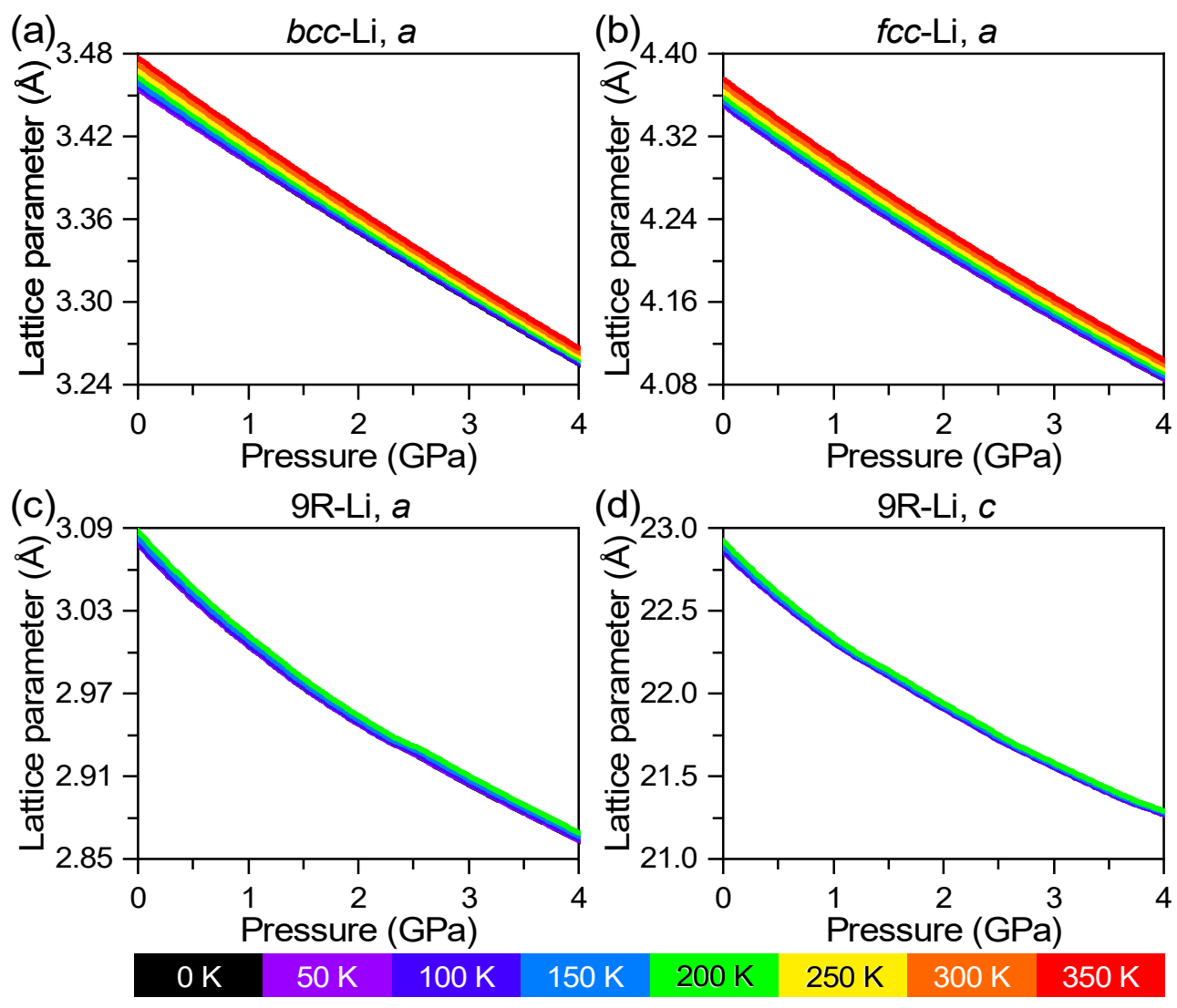

**Fig. S10** Temperature- and pressure-dependent equilibrium lattice parameters of (a) *bcc*-Li, (b) *fcc*-Li and (c-d) 9R-Li. Analogous results calculated at QHA level using the MACE force field are shown in Fig. S13.

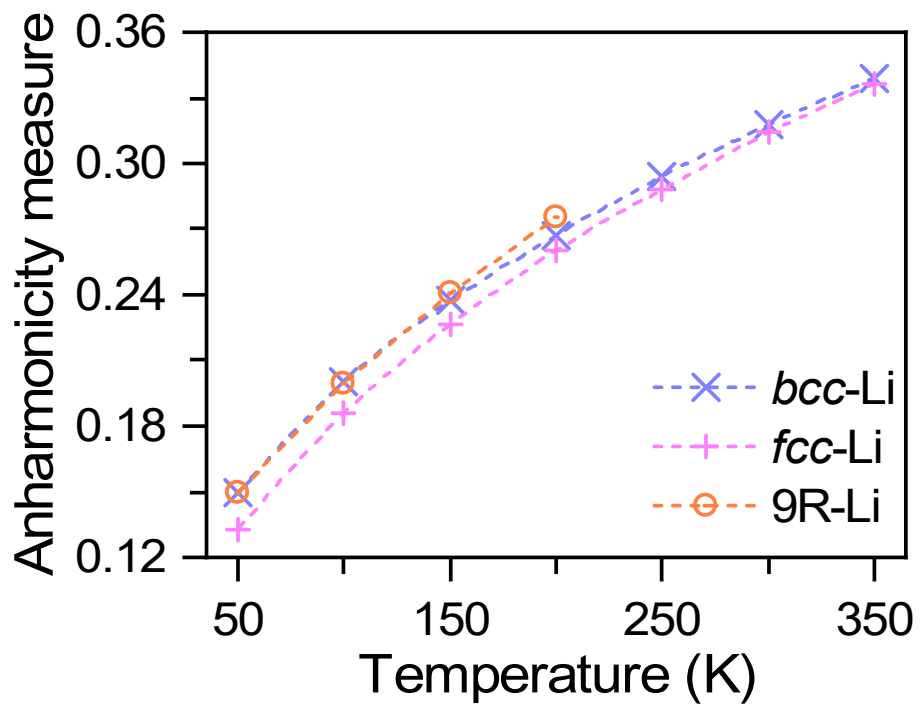

**Fig. S11** Analogous of Fig. 3, but excluding nuclear quantum effects.

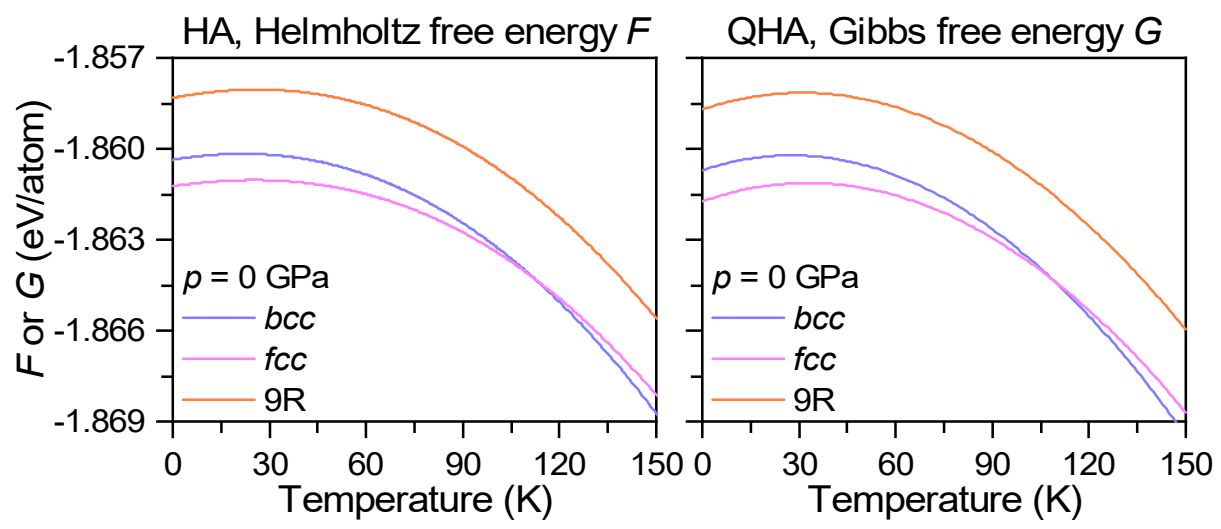

**Fig. S12** Free energy profile of *bcc*-, *fcc*- and 9R-Li calculated at HA and QHA level, with respective predicted phase-equilibrium temperature of 112 and 109 K.

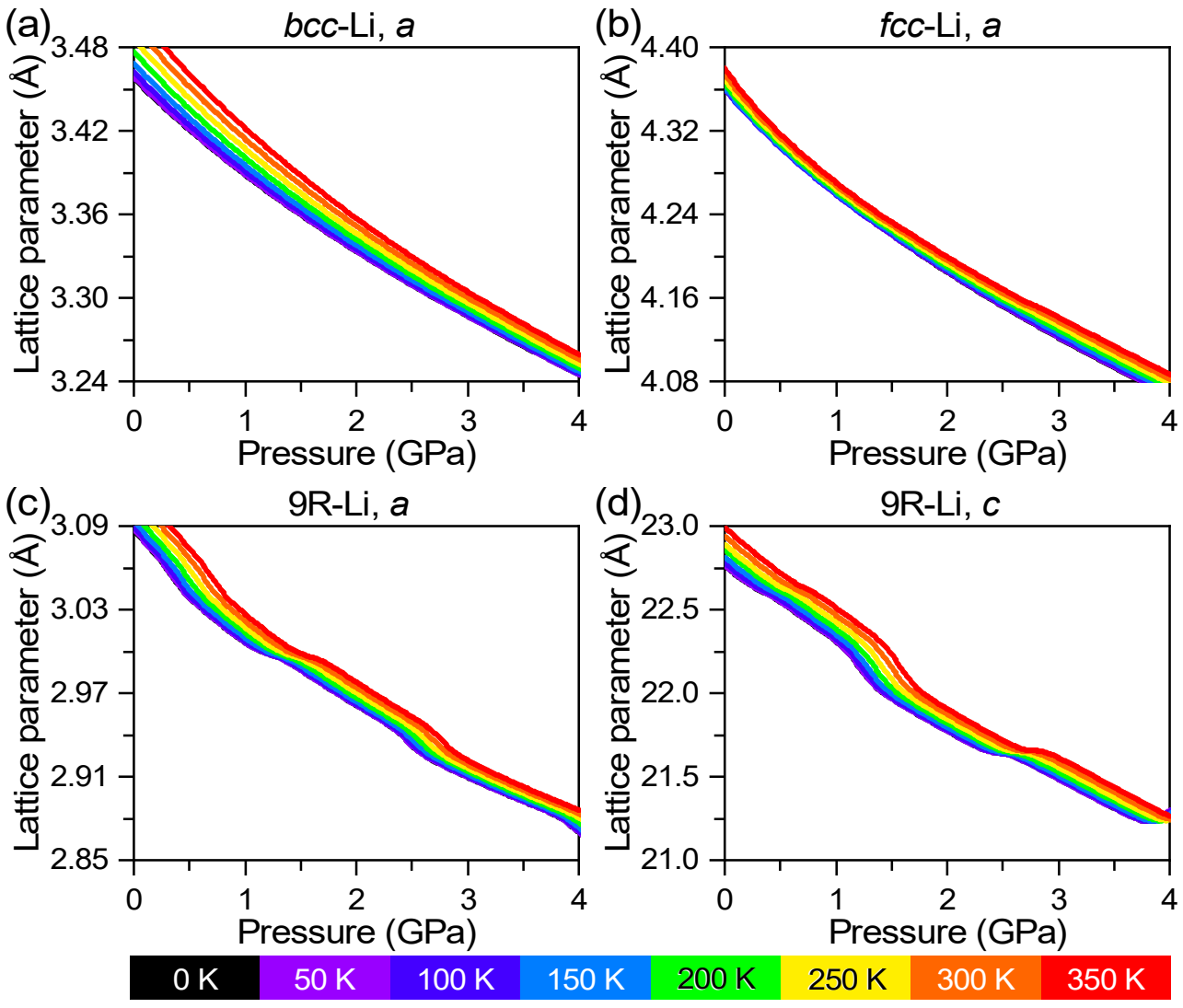

**Fig. S13** Analogous of Fig. S10, but calculated at QHA level using the MACE force field. Anomaly in equilibrium lattice parameters of 9R-Li around 2 GPa could indicate overfitted double-well polynomial from errored flat free energy profile around that section. In general, thermal expansion is more extensive as compared to results in Fig. S10.

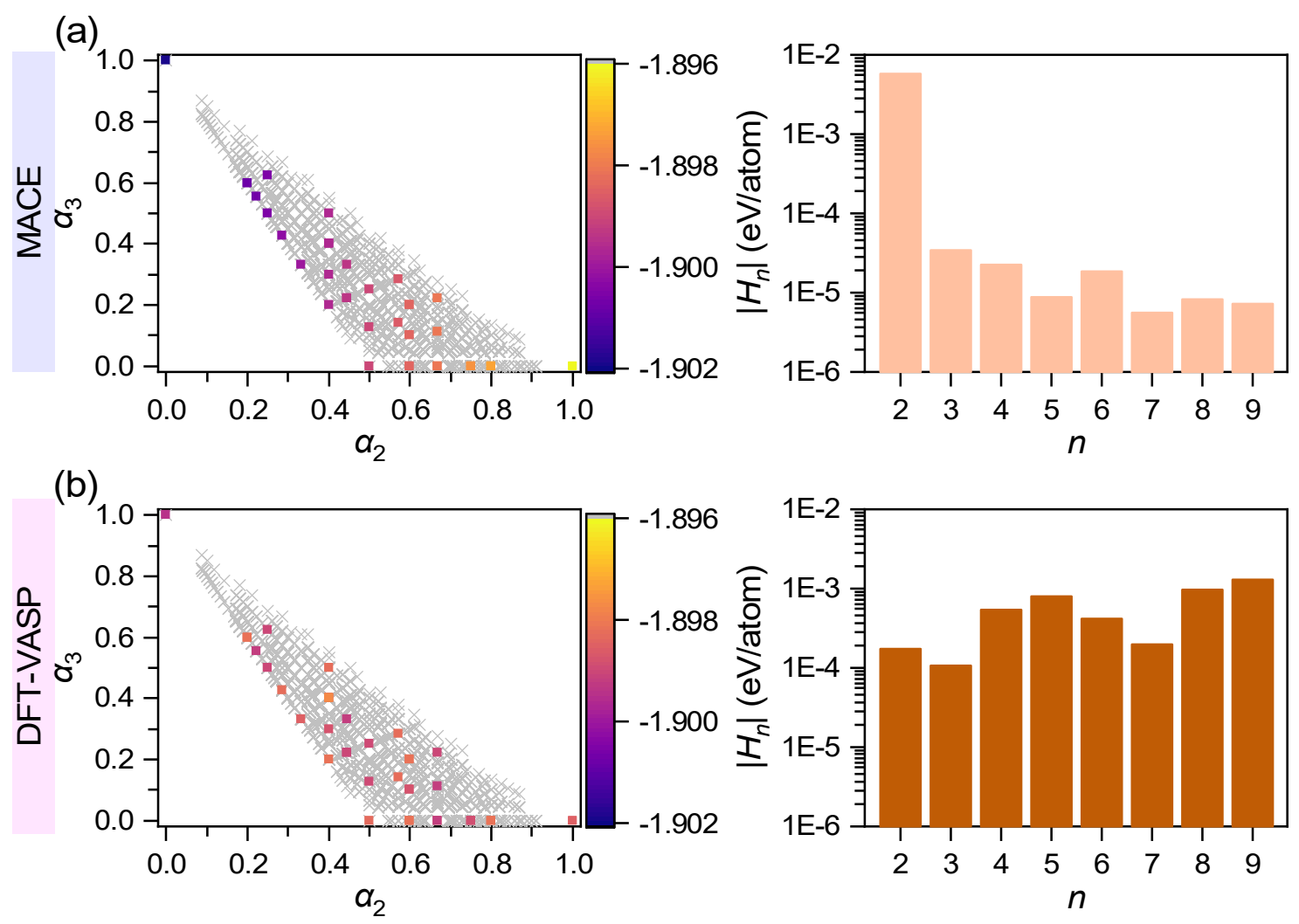

**Fig. S14** (a) Realizable close-packed stacking variants up to 22 layers per repeating unit, projected on the $\alpha_2$-$\alpha_3$ plane. Configurations of up to 10-layer stacking were optimized by the MACE force field and colored according to its energy. (b) Absolute value of fitted $H_n$ parameters in logarithmic coordinate.

## References

1. G. Kresse and J. Furthmuller, *Phys. Rev. B*, 1996, **54**, 11169-11186.
2. G. Kresse and J. Furthmuller, *Comput. Mater. Sci.*, 1996, **6**, 15-50.
3. P. E. Blöchl, *Phys. Rev. B*, 1994, **50**, 17953-17979.
4. J. P. Perdew, K. Burke and M. Ernzerhof, *Phys. Rev. Lett.*, 1996, **77**, 3865-3868.
5. J. P. Perdew, M. Ernzerhof and K. Burke, *J. Chem. Phys.*, 1996, **105**, 9982-9985.
6. W. S. Morgan, J. J. Jorgensen, B. C. Hess and G. L. W. Hart, *Comput. Mater. Sci.*, 2018, **153**, 424-430.
7. G. L. W. Hart, J. J. Jorgensen, W. S. Morgan and R. W. Forcade, *J. Phys. Commun.*, 2019, **3**, 065009.
8. M. K. Phuthi, A. M. Yao, S. Batzner, A. Musaelian, P. Guan, B. Kozinsky, E. D. Cubuk and V. Viswanathan, *ACS Omega*, 2024, **9**, 10904-10912.